\documentclass[12pt,preprint]{aastex}

\newcommand{\etal}{{et al. }}

\slugcomment{to appear in the ApJ Supplement, Vol. 154, Spitzer Special Issue, 
September 2004}

\shorttitle{Off-nuclear Starbursts in the Antennae}
\shortauthors{Wang et al.}

\begin{document}


\title{The Off-nuclear Starbursts In NGC 4038/4039 (The Antennae Galaxies)}


\author{Z. Wang\altaffilmark{1,4}, G. G. Fazio\altaffilmark{1}, 
M. L. N. Ashby\altaffilmark{1}, J. -S. Huang\altaffilmark{1}, 
M. A. Pahre\altaffilmark{1},
H. A. Smith\altaffilmark{1}, S. P. Willner\altaffilmark{1},
W. J. Forrest\altaffilmark{2}, J. L. Pipher\altaffilmark{2}
and J. A. Surace\altaffilmark{3}}


\altaffiltext{1}{Harvard-Smithsonian Center for Astrophysics,
    60 Garden Street, Cambridge, MA 02138-1516}
\altaffiltext{2}{Dept. of Physics and Astronomy,  University of Rochester,
                  Rochester, NY 14627-0171}
\altaffiltext{3}{Spitzer Science Center, Caltech, Pasadena, CA 91125}
\altaffiltext{4}{email: zwang@cfa.harvard.edu}


\begin{abstract}
Imaging of the Antennae galaxies (NGC 4038/4039) with the Infrared 
Array Camera (IRAC) aboard the Spitzer Space Telescope reveals large 
concentrations of star forming activity away from both nuclei of the 
two merging galaxies. These images confirm earlier findings based on
ISO data with lower angular resolution. The short wavelength emission 
shows numerous compact sources identified as stellar clusters. At the 
longer wavelengths, bright, more amorphous and filamentary features 
correlate well with the known distributions of denser gas, warm 
dust, and HII regions. There are also fainter, more diffuse components 
at all wavelengths that permeate the entire region and extend into 
the two tidal tails. Non-stellar dust emission dominates the 
5.8 and 8.0$\mu $m images, accounting for as much as 79\% of the light
at 5.8$\mu $m and 95\% at 8$\mu $m, averaged over the entire galaxy. 
Assuming that the non-stellar emission traces star formation,
the IRAC data provide a view into the total underlying star 
forming activities unaffected by obscuration. Using the flux 
ratio of non-stellar to stellar emission as a guide, we map the local 
star formation rate in the Antennae and compare that to similar 
measurements in both normal and infrared-luminous galaxies. 
This rate in the active regions is found to be as high as those 
seen in starburst and some ultra-luminous infrared galaxies on 
``per unit mass'' basis. The two galactic centers actually have lower 
star forming rates than the off-nuclear regions despite the presence 
of abundant dense gas and dust, suggesting that the latter is a 
necessary but not sufficient condition for on-going star formation.
\end{abstract}



\keywords{galaxies: individual (\objectname{NGC4038/4039}) -- 
galaxies: individual (\objectname{Antennae galaxies}) -- 
infrared: interstellar: continuum -- stars: formation}


\section{Introduction}

At an approximate distance of 21 Mpc\footnote{In this paper we assume a
value of Hubble constant H$_0$ = 70 km s$^{-1}$Mpc$^{-1}$.}, the Antennae 
galaxies provide a vivid illustration of a merging pair of spiral disks 
carrying a large amount of gas and dust (Toomre \& Toomre 1972; Hibbard 
\etal 2001). Attention to this system has increased over the last 
ten years since the debut of the Hubble images and the subsequent 
analysis that revealed thousands of young star clusters possibly being 
formed as part of the merging process (Whitmore \& Schweizer 1995; 
Whitmore \etal 1999). Theoretical models (Mihos \etal 1993, 
Mihos \& Hernquist 1996; Barnes 2002) have shown that the 
overall morphology, with its long-stretched tidal tails, can be 
accounted for via numerical simulations and that the behavior 
of gas and dust components are a consequence of the gravitational 
interaction. 

Infrared observations from space have opened an important line of inquiry 
on mergers, beginning with the IRAS All-sky Survey's discovery linking 
the ``ultra-luminous'' infrared galaxies (ULIRGs) with apparent 
interaction/merger morphology (Sanders and Mirabel 1996). Although not 
counted as an ULIRG by its absolute luminosity, the Antannae system
is nevertheless a typical case study involving both galaxy merger and 
active star formation. Of special interest are the Infrared Space 
Observatory (ISO) observations of NGC 4038/4039 that show a significant 
fraction of the mid-infrared emission may actually come from the 
``overlap region'' that lies between the two nuclei (Viguroux \etal 1996; 
Mirabel \etal 1998). Several recent analyses have attempted to
compare the multi-wavelength datasets of this galaxy pair in order 
to understand the detailed correlation of various components (Xu \etal 
2000; Zhang \etal 2001; Zezas \etal 2002; Kassin \etal 2003). However, 
the existing mid-infrared data have been limited by the relatively 
low angular resolution and sensitivity compared to other datasets.

The successful launch and early operations of the Spitzer Space Telescope 
(Werner \etal 2004) brought greatly enhanced capabilities to mid- and 
far-infrared observations. As part of the Spitzer Guaranteed Time Observing
(GTO) program, we have used the Infrared Array Camera (IRAC, Fazio \etal 2004) 
to perform broad-band imaging of the Antennae galaxies in four mid-infrared 
bands covering 3 to 10$\mu $m. This paper reports the first results of our 
observations with IRAC, which provide at least an order of magnitude 
improvement in sensitivity and a factor of 3 or higher in angular resolution 
compared to previous instruments.

\section{Observations and Data Analysis }


The Spitzer observations of the Antennae were performed on 24 December 
2003. There were a total of 20 individual pointing positions for each of 
the two IRAC fields-of-view (FOV). At each pointing position, exposures 
were taken in all four IRAC bands (3.6 and 5.8, or 4.5 and 8.0$\mu $m), 
with a 5-point Gaussian dithering pattern and 12-second frame time.
The area observed with all four IRAC bands exceeds $20'\times 16'$, 
covering the entire extent of the galaxy pair (in V band) including the 
two long tidal tails, with at least 160 sec total exposure time at every 
position.

Data analysis was carried out with both the standard SSC science 
data pipeline and the SAO software package SIP\footnote{SIP 
(the SAO IRAC Pipeline) was developed by the IRAC Instrument Team 
at the Smithsonian Astrophysical Observatory.}. Standard dark subtraction,
linearity correction, and flatfielding were used. After removing instrumental 
effects and cosmic ray hits, each of the four IRAC bands was flux-calibrated 
and mosaicked to a single image covering the two galaxies using the 
``drizzle'' method (Fruchter \& Hook 2002). Photometry of extended 
sources was performed on 
each band by fitting surface brightness isophotes and measuring
fluxes with appropriate aperture correction. Astrometry of each data 
frame was performed based on known positions of 2MASS point sources, 
with resulting accuracy of better than 0\farcs 3. The FWHMs of point 
sources in the four bands range between 1\farcs 4 and 1\farcs6. 





\section{The Antennae As Seen in the Four IRAC Bands}

The central part of the mosaicked IRAC data covering the main body of 
the Antennae is shown in the left panels of Figure~1. The bright 
mid-infrared emission is concentrated in three separate areas: 
around the two galactic nuclei, the outer ``spiral arms'' region of 
NGC 4038, and the ``overlap'' region (see Figure 2) as previously 
suggested by the ISO team (Mirabel \etal 1998). 
In each region, the emission is organized into
large irregularly shaped patches of about $10''$ ($1$ kpc) in
scale, with well-resolved, smaller-scale structures. In addition to 
these extended emission regions, there are numerous bright unresolved 
point sources and a diffuse component in all four bands 
(see contours in Fig.~3a) that permeates the main bodies 
of the two galaxies and extends to beyond the
central region covered by the frames in Fig.~1.

Although the overall morphologies revealed in the four bands show 
broad similarities, there are clear differences. The two galaxies' 
cores appear relatively brighter and more symmetric in the 3.6 and
 4.5$\mu $m band. (The center of NGC 4038 is nearly circular, while 
that of NGC 4039 roughly an oval). On the other hand, the morphology 
in the 8$\mu $m band (and to a lesser extent also at 5.8$\mu $m) 
appears more filamentary and has amorphous, irregular features
(see Fig.2), even though the effective angular resolution differs 
only slightly. Based on existing spectroscopic data, it is almost 
certain that emission at these wavelengthsis arises from the 
so-called aromatic features (usually attributed to PAHs) at 6.2, 
7.7 and 8.6$\mu $m (Rigopoulou \etal 1999; Tran \etal 2001). Indeed, 
IRAC observations of other late-type galaxies (Helou \etal 2004;
Pahre \etal 2004; Willner \etal 2004) also found excess dust emission
at these wavelengths. The aromatic feature emission is often
associated with the warm dust in or near giant HII regions in an 
ISM-rich environment, and these regions are nearly always sites of 
active star formation. 

The diffuse, extended emission also shows differences among 
the four IRAC bands. Overall, the longer wavelengths exhibit 
relatively weaker diffuse emission and fainter tidal tails, further
suggesting that the component responsible for the dust emission is 
more concentrated toward the galaxies' main bodies, where one finds 
denser gas and more warm dust.

\section{Comparison of IRAC Images with Existing Data}

The IRAC images can be directly compared with the mid-infrared 
(7, 9, and 15$\mu $m) data from ISO, ground-based near-infrared 
(K-band) imaging (Kassin \etal 2003; Brandl 2004; Martini 2004) and 
spectroscopy (Liang \etal 2001; Mengel \etal 2001), as well as the 
optical data, including those from the Hubble (Whitmore \etal 1995, 1999). 
The 3.6 and 4.5$\mu $m data trace each other well and generally match 
features seen in the ground-based near-infrared images. This confirms 
that the emission in the 2 to $5\mu $m range is mainly arising from 
the underlying, older stellar population with ordinary colors.
Both our data and the near-infrared imaging data show that the 
distribution of the young stellar clusters as seen in the HST 
images (Whitmore \etal 1999)
are far from uniform: they are congregated in groups of size 
scales of about one to a few kpc. On larger scales they tend to
concentrate toward the northen spiral arm, the overlap region,
and the two nuclei where one also finds on-going star formation.

IRAC data in the 5.8 and 8.0$\mu$m bands, on the other hand, are 
generally more consistent with the longer wavelengths ISO results.
More specifically, the new dataset also reveals details not seen by 
ISO, including asymmetric emission structures near the two galactic 
nuclei, the filamentary and loop-shaped features in many of the 
star-forming regions, and the fainter, extended diffuse component 
that permeates the entire system. The closest correlation, however,
turns out to be with 
the distributions of H${\alpha}$ emission and the dust features 
seen in absorption in the visible-wavelength image (Figures 2 and 3a). 
Some of the hollow (or ``ring-like'') features have blue 
stellar clusters in the center (Whitmore \etal 1999), indicative
of supernovae or stellar wind-related ``superbubble''-type structures 
in the interstellar medium. Away from the two galactic nuclei, the 
emission at 5.8 and 8.0$\mu $m also shows good correlation with 
the radio continuum images at 4 and 6 cm (Hummel \& van der Hulst 
1986; Neff \& Ulvestad 2000) and the CO (1-0) map of molecular 
gas (Stanford \etal 1990; Wilson \etal 2000).

\section{Off-Nuclei Star Formation as Measured by the Enhanced PAH Emission}

Emission features from heated dust are ubiquitous in late-type 
galaxies and are found to be greatly enhanced in many starburst 
and ULIRG galaxies. The flux ratio between the dust emission 
bands and the mid-infrared continuum has been proposed
as a quantitative measure of the relative 
level of starburst activities among infrared luminous galaxies 
(Genzel \etal 1998; Laurent \etal 1999, 2001; Tran \etal 2001). 
This approach appears to be broadly successful in categorizing the 
different types of objects and their power sources. However, one of the 
complications is that most existing measurements of infrared-luminous 
galaxies are based on global properties, so it is not clear to what 
extent they are contaminated by the additional non-thermal power 
sources such as AGNs. For example, an analysis of 41 ULIRGs finds poor 
correlation between the PAH feature to continuum flux ratio at 7.7$\mu $m 
and other starburst indicators (Farrah \etal 2003).

The special perspective provided by the Antennae is that the most 
active star formation is seen in well-resolved regions far 
away from the galaxies' nuclei, so the potential ``contamination'' 
by an AGN is minimized. Moreover, the two longer-wavelength 
IRAC bands cover three of the strongest PAH features in this spectral 
range (at 6.2, 7.7, and 8.6$\mu $m), thus simultaneous observation 
with all four IRAC bands allows a straightforward comparison 
of the fluxes with and without the contribution of the PAH bands 
over the entire galaxy at a high spatial resolution. In order 
to better assess the contribution of the PAH bands, we define the
``non-stellar'' (or dust) component of the emission to be the 
{\sl difference} between the measured flux and a ``stellar continuum'' 
flux. The latter was calculated for each of the four bands by 
normalizing a stellar spectral energy distribution (SED) 
\footnote{At these wavelengths, the results are insensitive to
the actual stellar type selected. For convenience we used
the Vega A0 V type in the fit.}
to an average of IRAC 3.6 and 4.5$\mu $m flux at each 
position. The right panel in Figure 1 illustrates the 
distribution of the non-stellar emission component. 

There is some residual emission
at 4.5$\mu $m, while the 3.6$\mu $m image is slightly negative 
(over-subtracted, mainly in the overlap region). 
The residual 4.5$\mu $m emission can have several possible origins: 
either these regions are extremely highly obscured (with A$_v > 50$), 
such that even at 3.6$\mu $m, the selective extinction is still 
significant; or there could be additional 4.5$\mu $m emission sources 
causing the SEDs to deviate from the stellar template (stars of 
different stellar types than we assumed, or additional dust emission 
features).
Nevertheless, the overall picture from the right panel of Fig.~1 is 
that the stellar continuum fit to IRAC 3.6 and 4.5$\mu $m data is 
reasonably good and that most of the emission ($>90$\%) 
in these two bands indeed comes from the stellar contribution. 

The situation is entirely different for the 5.8 and 8.0$\mu $m 
images: most of the emission in these two IRAC bands comes from the
non-stellar (i.e., warm dust) component. In fact, over the entire 
galaxy, the fractional contribution of the non-stellar flux component 
is $>$79\% for the 5.8$\mu $m band and $>$95\% for the 8$\mu $m 
band. Moreover, since the overlap region is not as bright in the 
continuum as the two galactic cores, this fraction is even higher 
there. That this non-stellar flux is dominated by the PAH emission is 
supported by the fact that the average [5.8] $-$ [8.0] surface brightness 
ratio for the non-stellar component is consistently about 1.8 mag
in all positions. This is very similar to the ratio found in a sample
of late-type galaxies based on the IRAC Mid-infrared Hubble Atlas program
(Pahre \etal 2004). Theoretical model calculations of dust bands 
predict a similar factor (Li and Draine 2001). We therefore conclude 
that the warm dust emission dominates fluxes measured in the 5 to 
8$\mu $m range virtually everywhere in the main body of the Antennae.

The strong PAH emission in this spectral range is not 
surprising. Previous data and IRAC observations of other late-type 
galaxies all suggest such a result, especially for regions of known 
star formation. However, what stands out in the Antennae is the 
degree to which the dust emission overwhelms the mid-infrared continuum. 
The ratio of PAH flux to stellar continuum found in the active 
regions of this galaxy is much higher (by a factor of at least 4 to 5)
than in other late-type galaxies (Pahre \etal 2004) and is comparable 
to those of many more distant starburst and ULIRG galaxies 
observed so far with IRAC 
(Wang \etal 2004). Moreover, unlike the majority of infrared-luminous
galaxies, in this case the most intensive PAH fluxes are clearly
not coming from the galactic centers but instead from regions far 
off the two nuclei (i.e., in the outer arms and the overlap region).

Fig.~3b shows the flux ratio of the non-stellar and stellar 
continuum for IRAC 8 $\mu $m data. This ratio is similar to but 
not the same as the ``PAH 7.7 (or 6.2)$\mu $m to continuum'' ratio 
used by several groups (Genzel \etal 1998; Laurent \etal 1999) to 
measure the starburst contribution of total power output in ULIRGs. 
(The difference is mainly in accounting for the thermal continuum of warm 
dust.) If the 3.6 to 4.5$\mu $m continuum is proportional to the mass 
of the underlying stellar population, and the non-stellar dust emission
is proportional to the current star formation rate, then this ratio 
provides an approximate measure of the ``star formation rate 
per unit stellar mass''. Not surprisingly, its value increases 
dramatically (to 10 -- 30) in the overlap region and in the outer 
spiral arms around NGC~4038. The ratio map in Fig.~3b can be 
compared with Fig.~3a, in which the IRAC $8\mu$m emission is 
overlaid (as contours) on the H${\alpha}$ image (greyscale) of 
the Antennae. This comparison again suggests that in 
off-nuclear regions where current star formation is taking place,
both the PAH and H${\alpha}$ emission are bright. (The
dust extinction appears to have a lesser effect on the H${\alpha}$ 
than broad-band optical images in off-nuclear regions).
These are the same regions where ISO found a large increase 
in mid-infrared continuum at 15 $\mu $m (Mirabel \etal 1998) and 
high excitation of ionized gas as measured by the [NeIII]/[NeII] 
line ratio. In contrast, this non-stellar vs. stellar flux
ratio has a much lower value (between 2 and 5) in both nuclei of 
NGC~4038 and NGC~4039, even though the extinction there is higher.

Despite their prominence in dust emission, the active star forming 
regions involve only a relatively small amount of stellar mass. 
Assuming that the integrated 
luminosity in the shorter-wavelength IRAC bands is proportional to 
the stellar mass everywhere in the Antennae, the active 
star forming regions in Fig.~3b (defined as having the flux ratio 
of (non-stellar) 8.0 $\mu $m to stellar continuum greater than 7:1) 
comprise only about 10\% of the total stellar mass of the entire 
system (assuming a constant mass-to-light ratio while accounting 
for extinction based on the infrared data). 
The current bolometric luminosity of the Antennae is estimated
to be $4\times 10^{10}L_{\odot}$. But if we assume that a sufficient 
dense gas exists and that the amount of star formation per 
unit stellar mass remains as high as in the overlap region after 
the two galaxies finally merge into one, then the Antennae's
bolometric luminosity could be elevated to at least 10 times greater 
as a result of the merger, pushing it into the league of ULIRG 
galaxies such as UGC~5101 or Arp~220. It is at least theoretically 
possible that such circumstances could occur in the Antennae at 
a different stage of its merger from the present one.

\section{Discussion}

Several previous works have postulated that the ultimate regulator 
for the level of star forming activities in a starburst galaxy is the 
available supply of raw materials, i.e., the amount of cold, dense 
molecular gas (Efstathiou \& Rowan-Robinson 1995; Efstathiou \etal 
2000; Gao \etal 2001). Our analysis of IRAC data on the Antennae 
only partially supports this idea: the areas of star forming activities 
very roughly coincide with the CO intensity maps (Wilson \etal 2000, 2003)
in the off-nuclear regions. However, the amount of star formation per 
unit stellar mass is considerably
lower in the two galaxies' nuclear regions where, despite
the presence of plenty of molecular gas, neither H${\alpha}$ nor 
dust emission is proportionally brighter. In other words, the mere 
abundance of dense gas is a necessary but may not be a sufficient 
condition for local starbursts in interacting galaxies. Other factors,
such as shocks, are also likely to be important at least in the stages
prior to the final coalescence.

Using the flux ratio of mid-infrared dust emission vs. stellar continuum 
as an indicator of star forming activities may be more appropriate for 
the Antennae because most of star formation is off-nuclear and thus the 
effect of AGN is minimal. Our results suggest that the star forming 
regions in this galaxy are as active as those in the more luminous 
infrared galaxies, but the activity is localized rather than global. 
It remains to be addressed whether the nuclear regions of the Antennae 
have passed their peak star forming epoch or a more active starburst 
will take place as the merger evolves.






\acknowledgments

We thank Dr. Bradley C. Whitmore for his help in providing digitized HST data 
for comparison. Discussions with Drs. Pauline Barmby, Joseph L. Hora, 
Messrs. Daisuke Iono and Junzhi Wang have helped to improve the manuscript. 
This work is based on observations made with the Spitzer Space Telescope, 
which is operated by the Jet Propulsion Laboratory, California Institute 
of Technology under NASA contract 1407. Support for this work
was provided by NASA through Contract Number 1256790 issued by JPL/Caltech. 
Support for the IRAC instrument was provided by NASA through Contract 
Number 960541 issued by JPL.



Facilities: \facility{Spitzer(IRAC)}.





\clearpage



\begin{figure}
\vskip 5.0truein
\centerline{(Figure 1)}
\vskip 1.0truein
\caption{Left panel: greyscale plot of the central part of the 
Antennae in the four IRAC bands: from top to bottom: 3.6, 4.5, 5.8 
and 8.0$\mu $m bands. Right panel: same as the left, but for the 
non-stellar emission only. The continuum emission, generated by 
extrapolating the average of the 3.6 and 4.5 $\mu $m images to the 
appropriate wavelength band, is subtracted. North is up and East 
is to the left. The angular size of each image is about 4\farcm 3 by
3\farcm 1 in RA and Dec directions, respectively.
The sky background in the 8 $\mu $m image appear smoother 
than the other bands because the surface brightness of 
the emission is much higher.
\label{fig1}}
\end{figure}


\begin{figure}

\vskip 5.0truein
\centerline{(Figure 2)}
\vskip 1.0truein

\caption{(Color Plate 1.) Left: Color-composite image of the 
central part of the Antennae based on data from the optical 
(U, V) bands (Kuchinski \etal 2000) and the IRAC 8$\mu $m band: 
Blue: U, Green: V, Red: 8$\mu $m. Right: the HST UBV and 
H${\alpha}$ color composite for comparison. The two insets on the 
upper right show expanded views of a star forming region in the 
northern spiral arms region. The HST data were taken with the WFPC
camera and adapted from Whitmore \etal (1999). The orientation in 
this figure is indicated by the arrow in the right panel and is
different from Fig.~1 and Fig.~3. \label{fig2}}
\end{figure}


\begin{figure}
\vskip 5.0truein
\centerline{(Figure 3)}
\vskip 1.0truein
\caption{(Color Plate 2.) [$a$] (left): The IRAC 8$\mu$m emission (as contours) overlaid on the 
H$\alpha$ image of the same region (as grey scales). The H$\alpha$ image is from 
the WFPC camera of the Hubble Space Telescope (Whitmore \etal 1999). The contours
are from 5 to 120 in steps of 10, in the IRAC flux unit of 
MJy/sr. The outer contour shows the extended diffuse component at 8$\mu$m. \/
[$b$] (right): Ratio of the 8.0$\mu $m non-stellar flux to 4.5$\mu$m stellar 
continuum, shown in both greyscale and contours. The contour levels are 
from 10 to 35, in steps of 5.
 \label{fig3}}
\end{figure}









\clearpage




\end{document}